\documentclass[12pt,a4paper]{iopart}

\usepackage{graphicx}
\usepackage{color}
\usepackage{mathrsfs} 
\usepackage{dsfont}
\usepackage{iopams}  

\usepackage[bookmarks=true,
            bookmarksopen=true,colorlinks=false]{hyperref}

\newcommand{\eqref}{\eref}
\newcommand{\dd}{\mathrm{d}}
\newcommand{\ee}{\mathrm{e}}

\newcommand{\R}{\mathds R}

\newcommand{\eps}{\varepsilon}

\newcommand{\Scri}{\mathscr{I}}
\newcommand{\JH}[1]{{\color{red}#1}}

\renewcommand{\JH}[1]{#1}

\begin{document}

\title[Fully pseudospectral solution of the conformally invariant wave equation III.]{Fully pseudospectral solution of the conformally invariant wave equation near the cylinder at spacelike infinity. III: Nonspherical Schwarzschild waves and singularities at null infinity}

\author{J\"org Frauendiener and J\"org Hennig}
\address{Department of Mathematics and Statistics,
           University of Otago,
           PO Box 56, Dunedin 9054, New Zealand}
\eads{\mailto{joergf@maths.otago.ac.nz} and \mailto{jhennig@maths.otago.ac.nz}}

\begin{abstract}
We extend earlier numerical and analytical considerations of the conformally invariant wave equation on a Schwarzschild background from the case of spherically symmetric solutions, discussed  in Class.\ Quantum Grav.\  {\bf 34}, 045005 (2017), to the case of general, nonsymmetric solutions. A key element of our approach is the modern standard representation of spacelike infinity as a cylinder. With a decomposition into spherical harmonics, we reduce the four-dimensional wave equation to a family of two-dimensional equations. These equations can be used to study the behaviour at the cylinder, where the solutions turn out to have, in general, logarithmic singularities at infinitely many orders. We derive regularity conditions that may be imposed on the initial data, in order to avoid the first singular terms. We then demonstrate that the fully pseudospectral time evolution scheme can be applied to this problem leading to a highly accurate numerical reconstruction of the nonsymmetric solutions. We are particularly interested in the behaviour of the solutions at future null infinity, and we numerically show that the singularities spread to null infinity from the critical  set, where the cylinder approaches null infinity. The observed numerical behaviour is consistent with similar logarithmic singularities found analytically on the critical set. Finally, we demonstrate that even solutions with singularities at low orders can be obtained with high accuracy by virtue of a coordinate transformation that converts solutions with logarithmic singularities into smooth solutions.
\\[2ex]{}
\end{abstract}


\section{Introduction\label{sec:intro}}

The present paper continues the study of the asymptotic behaviour of solutions to the conformally invariant wave equation on a Schwarzschild background using analytical and numerical means.  The motivation for this and the previous two articles~\cite{FrauendienerHennig2014,FrauendienerHennig2017} is the desire to better understand the interplay between smoothness of the solutions at null infinity and certain fall-off conditions on their initial data on a Cauchy surface. This is a fascinating property of conformally invariant equations on asymptotically flat spacetimes, which was \JH{discovered by} Friedrich (see~\cite{Friedrich2004} and references therein) in the case of the conformal field equations for gravitational fields.

The situation may be briefly summarised as follows. The idea of using conformal methods for the analysis of asymptotic properties of the gravitational field goes back to Penrose who, in 1965~\cite{Penrose1965}, introduced the concept of asymptotically simple spacetimes as those which admit the attachment of a conformal boundary. In the case of the Minkowski spacetime, the asymptotic boundary turns out to consist of two null hypersurfaces, the past and future null infinities $\Scri^\pm$. The future cone emanates from a regular point, spacelike infinity $i^0$, and focuses \JH{to} another point, future timelike infinity $i^+$. The past cone behaves similarly, emanating from the past timelike infinity $i^-$ and focusing at $i^0$. Asymptotically flat nonvacuum solutions with vanishing cosmological constant show almost the same structure of two null hypersurfaces as the conformal boundary, except that the ``points'' $i^0$ and $i^\pm$ are no longer regular points. This is not surprising in the case of $i^\pm$, since all the matter in the spacetime must have come out of $i^-$ and will end up in $i^+$. Instead, the singularity of $i^0$ is due to the gravitational field itself. Even in the case of the Schwarzschild spacetime, which does not contain any matter but features a nonvanishing total ADM mass, the ``point'' $i^0$ is singular.

The structure of the conformal boundary suggests that the late incoming radiation and the early outgoing radiation somehow communicate via $i^0$. For a long time, the exact structure of the spacetime region near $i^0$ remained unknown. Several authors tried to capture the essential behaviour by imposing appropriate conditions on the gravitational fields or the metric. But these conditions were to some extent \emph{ad hoc} and did not make full use of the field equations. In~\cite{Friedrich1998}, Friedrich succeeded in shedding more light on spacelike infinity by using all of the available conformal freedom for the Einstein equations. In particular, he used Weyl connections and their associated geodesics to introduce a conformally invariant gauge for the field equations.

When the spacetime structure is analysed using the field equations in this gauge, the previously unstructured singular ``point'' $i^0$ emerges as an entire cylinder $I$ of topology $S^2 \times \R$ meeting $\Scri^\pm$ in a 2-sphere labelled $I^\pm$. An asymptotically Euclidean hypersurface meets the cylinder in a 2-sphere $I^0$, distinct from $I^\pm$, and initial data for the Einstein evolution induce data for the field variables on $I^0$.

We may think of the cylinder as the bridge across which information travels from $\Scri^-$ to $\Scri^+$. This viewpoint is corroborated by the fact that, on the cylinder, the field equations reduce to an entirely intrinsic system of equations. Thus, the cylinder $I$, which is a boundary to the physical manifold, is not a boundary for the field equations in the sense that there is no place to impose boundary conditions on $I$. Indeed, the full set of derivatives of the field variables normal to $I$ is determined completely by an intrinsic symmetric hyperbolic system of transport equations and initial data on $I^0$.

The previously singular nature of $i^0$ is reflected in the fact that the transport equations for the normal derivatives degenerate on the spheres $I^\pm$, where the cylinder meets the null infinities. As a consequence, their solutions develop logarithmic singularities at $I^\pm$ unless their initial data on $I^0$ are restricted appropriately. The structure of the regularity conditions is hierarchical in the sense that the logarithmic divergence in a certain order of the normal derivatives of the field variables can be eliminated provided that they have been eliminated at all lower orders.
Thus, logarithmic divergences up to a certain order can be avoided by suitable change of the initial data subject to fall-off conditions in the physical spacetime.

When the regularity conditions on $I^0$ are violated, there will be singularities at $I^+$. Since $\Scri^+$ is causally in the domain of influence of $I^+$, the hyperbolic nature of the evolution equations suggests that the singularities will also propagate to null infinity and diminish its smoothness, thus spoiling, in particular, the peeling properties. It is no surprise that a detailed discussion of this behaviour is rather complicated (see the work of Valiente Kroon~\cite{ValienteKroon2012} and references therein) and hence, until now, only some special cases and analogous systems have been analysed in \JH{complete} detail. 

Since the degeneracies occur in the part of the conformal field equations addressing the Weyl curvature, namely the Bianchi equations, it is natural to investigate the linearised Bianchi equations. Friedrich~\cite{Friedrich2003} discussed this system on a Minkowski background from the point of view of the well-posedness of the initial value problem. He found estimates for the solutions in terms of the initial data, which show that solutions of these spin-2 zero-rest-mass equations lose smoothness on $\Scri^+$ depending on how many conditions in the hierarchy of regularity conditions have been satisfied.

Related analytical investigations of the Maxwell equations on a Schwarzschild background were carried out by Valiente Kroon in \cite{ValienteKroon2007,ValienteKroon2008}. It was shown that the electromagnetic field generally contains logarithmic singularities at $I^\pm$, unless the initial data satisfy additional regularity conditions. Moreover, detailed estimates of integrals of the field components and their derivatives in terms of the initial data were derived.

In~\cite{FrauendienerHennig2014} we analysed the behaviour of the conformally invariant wave equation, i.e.\ the zero-rest-mass equation for spin zero, on a Minkowski background,  with spatial infinity represented as the cylinder $I$, in order to find out to what extent the behaviour of the conformal field equations is shared by this simple system. We found that the solutions also need to satisfy certain conditions on $I^0$ to prevent the appearance of logarithmic singularities at $I^+$ (and $I^-$). In~\cite{FrauendienerHennig2017} we generalised this to the conformally invariant wave equation on the Schwarzschild spacetime finding the same general structure, albeit somewhat more complicated: in the Minkowski case, global regularity can be achieved if just a single regularity condition is imposed, which eliminates all singularities. However, in the Schwarzschild case, only rather trivial solutions are regular everywhere. 

In both papers \cite{FrauendienerHennig2014,FrauendienerHennig2017}, we restricted ourselves to spherical symmetry to simplify the calculations, and we used fully pseudospectral methods to compute global solutions. Moreover, with our previous investigations in the Schwarzschild case, we could only show existence of logarithmic singularities at $I^+$, but
we were not able to give evidence of the loss of smoothness of the solutions on $\Scri^+$.
Hence, with the present paper, we pursue two aims: the first is to generalise our study to the \JH{case} without any symmetry, and the second is to use the very direct relationship between the smoothness of functions and \JH{the} convergence properties of their spectral approximations to give clear evidence about the loss of smoothness on $\Scri^+$ of the solutions depending on how many regularity conditions on $I^0$ have been imposed.

Our main tool for the numerical investigations in this paper is the fully pseudospectral method introduced in \cite{HennigAnsorg2009}, which was first applied to several physical problems in \cite{AnsorgHennig2011} and \cite{Hennig2013}. The basic idea is to approximate the unknown functions using Chebyshev interpolation, and to derive and solve an algebraic system of equations for the Chebyshev coefficients. The algebraic equations are obtained from the requirement that the wave equation or appropriate boundary and regularity conditions are satisfied at a grid of  collocation points. Details can be found in the above references and in \cite{FrauendienerHennig2014, FrauendienerHennig2017}.
The most important advantages of this numerical approach are the extremely accurate solutions that one can typically obtain (usually close to machine accuracy), and the highly-implicit nature of the solution process that effectively couples all gridpoints. A convenient consequence of the latter property is that the Courant-Friedrichs-Lewy (CFL) condition, which restricts the size of the time stepsize in traditional time-marching schemes, is not required here.\footnote{As a related example where the CFL condition creates difficulties, we mention the numerical studies of the spin-2 system on a Minkowski background in \cite{BeyerDoulis2012,BeyerDoulis2014,DoulisFrauendiener2013}. With the numerical method used in these references, it was found that future null infinity $\Scri^+$ can only be asymptotically approached with a \JH{very large} number of steps. Indeed, the divergence of the characteristic speeds at $\Scri^+$ enforces, via the CFL condition, \JH{smaller} and smaller time steps. With the fully pseudospectral scheme, on the other hand, we can include $\Scri^+$ \JH{in} the numerical domain and directly reach it \JH{using only} a moderate number of grid points.}

As discussed above, a very important question that was not addressed in \cite{FrauendienerHennig2017} is whether the logarithmic singularities are restricted to the set $I^+$ or whether they are present on $\Scri^+$ as well. Here we investigate this question with a careful analysis of the numerical solutions, and we show that singularities indeed also appear everywhere on~$\Scri^+$. 

A particular new feature that we introduce for our numerical considerations in this article is an improved numerical reconstruction of more singular solutions. To this end, we describe a modification of the numerical method, based on a reformulation of the problem in adapted coordinates. This will allow us to resolve solutions with singularities at lower orders with much greater accuracy. 

This paper is organised as follows. In Sec.~\ref{sec:coords}, we derive the conformally invariant wave equation and the corresponding decomposition into spherical harmonics in a suitable conformal representation of the Schwarzschild spacetime. \JH{In} Sec.~\ref{sec:cylinder}, we derive analytical properties of the solutions near the cylinder.
In order to test applicability of the pseudospectral scheme to the present problem, we derive simple exact solutions in Sec.~\ref{sec:testsol} that are completely free of singularities,  and we reconstruct them numerically. \JH{In Sec.~\ref{sec:logsing}}, we extend the numerical considerations to singular solutions and localise singularities on $\Scri^+$. \JH{The reformulation} of the problem for more accurate numerical computations  \JH{is} described in Sec.~\ref{sec:reg}. Finally, in Sec.~\ref{sec:discussion}, we summarise and discuss our results.

\section{Coordinates and wave equation\label{sec:coords}}

We study the conformally invariant wave equation for a scalar function $f$, i.e.\ the equation
\begin{equation}\label{eq:CWE}
 g^{ab}\nabla_a\nabla_b f-\frac{R}{6}f=0.
\end{equation}
Here, $g$ is a fixed background metric, which we take to be conformal to the Schwarzschild metric, and $R$ is the corresponding scalar curvature. We want to consider this equation near the cylinder representation $I$ of spacelike infinity $i^0$ and near a portion of future null infinity $\Scri^+$. To this end,  we use the same coordinates $(\tau,\rho,\theta,\phi)$, first introduced by Friedrich \cite{Friedrich2004}, that we also utilised in \cite{FrauendienerHennig2017}. In terms of these coordinates, the conformal metric takes the form
\begin{equation}\label{eq:conmet}
 g=\frac{2}{\rho}A\,\dd\rho\,\dd\tau-\frac{1-\tau}{\rho^2}A[2-(1-\tau)A]\dd\rho^2-\dd\theta^2-\sin^2\theta\,\dd\phi^2,
\end{equation}
where
\begin{equation}\label{eq:defA}
 A:=\frac{(1-r)(1+\rho)^3}{(1-\rho)(1+r)^3},\quad r:=\rho(1-\tau).
\end{equation}
This is related to the physical metric $\tilde g$ of the Schwarzschild spacetime with mass parameter $M$ via $g=\Theta^2\tilde g$ with conformal factor
\begin{equation}\label{eq:Theta}
 \Theta=\frac{2r}{M(1+r)^2}.
\end{equation}
We consider the coordinates in the intervals $0\le\rho\le\rho_\mathrm{max}<1$ and $0\le\tau\le1$, where $\rho=0$ corresponds to the (future half of the) cylinder $I$ and $\tau=1$ to a part of $\Scri^+$. The coordinates are adapted to one family of radial null geodesics, which take the simple form $\rho=\mathrm{constant}$. Note that these coordinates only cover a part of the extended Schwarzschild spacetime, and, in particular, do not extend to the event horizon, cf.~\cite{FrauendienerHennig2017} for more details. Here, however, we are mainly interested in a vicinity of infinity. Nevertheless, if required, the coordinates can be modified to cover a larger region including the horizon \cite{FrauendienerHennig2017}.

In terms of the background metric \eqref{eq:conmet}, the conformally invariant wave equation for a function $f=f(\tau,\rho,\theta,\phi)$ reads
\begin{equation}\fl
 (1-\tau)[2-(1-\tau)A] f_{,\tau\tau}+2\rho f_{,\tau\rho}
 -2\left[1-(1-\tau)A\frac{1-2r}{1-r^2}\right]f_{,\tau}
 -A\bigtriangleup_\Omega f+\frac{4rA}{(1+r)^2}f=0,
\end{equation}
where $\bigtriangleup_\Omega f=\frac{1}{\sin\theta}(\sin\theta\,f_{,\theta})_{,\theta}-\frac{1}{\sin^2\theta} f_{,\phi\phi}$ is the angular part of the flat Laplacian. Since the $\theta$- and $\phi$-derivatives only appear in the form of $\bigtriangleup_\Omega$, we can decompose the solutions into spherical harmonics $Y_{lm}$ in the usual way,
\begin{equation}
 f(\tau,\rho,\theta,\phi)=\sum_{l=0}^\infty\sum_{m=-l}^l\psi_{lm}(\tau,\rho)Y_{lm}(\theta,\phi).
\end{equation}
This reduces the wave equation to a set of independent equations for the modes $\psi_{lm}$,
\begin{eqnarray}\label{eq:modes}
 (1-\tau)[2-(1-\tau)A] \psi_{lm,\tau\tau}+2\rho \psi_{lm,\tau\rho}
 -2\left[1-(1-\tau)A\frac{1-2r}{1-r^2}\right]\psi_{lm,\tau} \nonumber\\
 \quad +Al(l+1) \psi_{lm}+\frac{4rA}{(1+r)^2}\psi_{lm}=0.
\end{eqnarray}
Note that only $l$ but not $m$ appears in the equations. Effectively, we can therefore study $1+1$ dimensional problems for each $l$. For ease of notation, we will usually assume that we have made a fixed choice of $l$ (and $m$) and denote the corresponding mode simply by $\psi$.

\section{Behaviour near the cylinder\label{sec:cylinder}}

In the spherically symmetric case, discussed in \cite{FrauendienerHennig2017}, we found that the wave function and its spatial derivatives at the cylinder $\rho=0$ generally have logarithmic terms. As a consequence, certain mixed derivatives diverge as $\tau\to1$, i.e.\ at the set\footnote{Note that $I^+$ is a topological sphere, but in two dimensional diagrams that suppress the angular coordinates, it is represented by a point.} $I^+$. The order at which the first singularities appear depends on the choice of initial data, and we can systematically construct data for which the lowest singularities up to a certain order are all eliminated. In the following we show that these results carry over to the nonspherical case.

We consider a solution $\psi(\tau,\rho)$ that can be expanded in a Taylor series about $\rho=0$ with $\tau$-dependent coefficients, 
\begin{equation}\label{eq:expan}
 \psi(\tau,\rho)=\psi_0(\tau)+\rho\psi_1(\tau)+\rho^2\psi_2(\tau)+\dots
\end{equation}
Plugging this into \eqref{eq:modes}, we obtain a system of ODEs of the form
\begin{equation}\label{eq:ODE}
 (1-\tau^2)\ddot\psi_n+2(n-\tau)\dot\psi_n+l(l+1)\psi_n=R_n,
\end{equation}
where the dot denotes differentiation with respect to $\tau$.
The inhomogeneity $R_n$ on the right-hand side depends linearly on up to six previous functions $\psi_{n-1}$, $\psi_{n-2},\dots$ and their first and second order $\tau$-derivatives. While $R_n$ can be given explicitly as a lengthy expression, we can study some properties of the solutions without using this explicit form.

In the spherically symmetric case $l=0$, we can easily integrate \eqref{eq:ODE} directly. For $l>0$, we can solve \eqref{eq:ODE} by observing that an appropriate rescaling of $\psi_n$ transforms the equation into an inhomogeneous associated Legendre equation. It turns out that  we have to dicsuss the two cases $n\le l$ and $n>l$ separately.

We begin with the case $n=0,1,\dots,l$. If we abbreviate the associated Legendre functions of the first and second kind with $u=P^n_l(\tau)$ and $v=Q^n_l(\tau)$, then the general solution to \eqref{eq:ODE} can be written as
\begin{eqnarray}\label{eq:ODEsol1}\
 \psi_n(\tau)&=&\frac{(l-n)!}{(l+n)!} \alpha_n(\tau)
 \left[\left(-\int_0^\tau\frac{R_n(\tau') v(\tau')}{\alpha_n(\tau')}\,\dd\tau'+c_1\right)u(\tau)\right.\nonumber\\
 &&\qquad
      +\left.\left(\int_0^\tau\frac{R_n(\tau') u(\tau')}{\alpha_n(\tau')}\,\dd\tau'+c_2\right)v(\tau)\right]
\end{eqnarray}
in terms of two integration constants $c_1$, $c_2$ and with 
\begin{equation}
 \alpha_n(\tau):=\left(\frac{1-\tau}{1+\tau}\right)^{\frac n2}.
\end{equation}
From this result we can now derive the following. If we assume that, for fixed $l$, $R_n$ is regular (namely if singularities at previous orders $n-1$, $n-2,\dots$ have all been eliminated), then it is possible to choose a specific value for $c_2$ (and an arbitrary value for $c_1$) such that the solution $\psi_n$ is regular as well. More explicitly, using certain properties of associated Legendre functions, the solution \eqref{eq:ODEsol1} can be reformulated as
\begin{eqnarray}\label{eq:ODEsol2}
 \psi_n &=& \frac{(l-n)!}{2(l+n)!}P^{(n)}_l
  \Bigg[\sum_{k=0}^{n-1}\Bigg(\begin{array}{c}n\\k\end{array}\Bigg)
   (-1)^{n-k}\frac{\dd^{n-k-1}}{\dd\tau^{n-k-1}}\left(\tilde R P^{(k)}_l\right)\Big|_{\tau=1}\nonumber\\
  && +\int_0^1\tilde R P^{(n)}_l\dd\tau-(-1)^n c_2\Bigg](1-\tau)^n\ln(1-\tau)
  +\textrm{regular terms},
\end{eqnarray}
where we abbreviated the $k$th derivative of the Legendre polynomial $P_l$ with $P^{(k)}_l$, and where we defined $\tilde R:=(1+\tau)^n R_n$. Hence we observe that, if previous orders are free of singularities, then the function $\psi_n$ generally contains a logarithmic term of the form $(1-\tau)^n\ln(1-\tau)$. However, we can always choose $c_2$ such that the square bracket in \eqref{eq:ODEsol2} vanishes, corresponding to a particular choice of initial data. In that case $\psi_n$ is regular too. Therefore, by imposing more and more constraints onto the initial data, we can step-by-step eliminate more and more singular terms.

Now we look at the second case $n=l+1,l+2,\dots$ For $n>l$, the above function $u=P_l^n(\tau)$ vanishes identically, because it is proportional to the $(n+l)$th derivative of a polynomial of degree $2l$, namely to $\frac{\dd^{n+l}}{\dd\tau^{n+1}}(1-\tau^2)^l$. Consequently, Eq.~\eqref{eq:ODEsol1} does not provide the general solution in that case. We find, however, that the replacement
\begin{equation}
 \frac{\dd^{n+l}}{\dd\tau^{n+l}}(1-\tau^2)^l\to
 \lim_{\eps\to0}\frac{1}{\eps}\frac{\dd^{n+l}}{\dd\tau^{n+l}}(1-\tau^2)^{l+\eps}
 =\frac{\dd^{n+l}}{\dd\tau^{n+l}}(1-\tau^2)^l\ln(1-\tau^2)
\end{equation}
does not vanish for $n>l$, but still leads to a solution. Based on this observation, we can construct the general solution in terms of the new functions
\begin{equation}
 u:=(1-\tau)^n\frac{\dd^n}{\dd\tau^n}[P_l\ln(1-\tau^2)],\quad
 v:=(1-\tau)^n\frac{\dd^n}{\dd\tau^n}\left[P_l\ln\frac{1-\tau}{1+\tau}\right].
\end{equation}
Note that the Wronskian $W$ of $u$ and $v$ turns out to be of the form 
$W\equiv u\dot v-\dot u v=c(1-\tau)^{n-1}/(1+\tau)^{n+1}$. In terms of the constant $c$ in this formula, the general solution can be written as
\begin{equation}\label{eq:ODEsol3}
 \fl
 \psi_n=\frac{1}{c}\left[\left(-\int_0^\tau\left(\frac{1+\tau'}{1-\tau'}\right)^nR_n v\,\dd\tau'+c_1\right)u
 +\left(\int_0^\tau\left(\frac{1+\tau'}{1-\tau'}\right)^nR_n u\,\dd\tau'+c_2\right)v\right].
\end{equation}
With a similar reformulation as above, we find that, for regular $R_n$, the particular solution with $c_1=c_2=0$ generally behaves like $(1-\tau)^n\ln(1-\tau)$ near $\tau=1$. Here, however, it is not possible to choose $c_1$ and $c_2$ in the general solution such that this singularity is eliminated, because $u$ and $v$, and, therefore, any linear combinations $c_1u+c_2v$ are regular functions for $n>l$. Consequently, they cannot compensate for singularities in the integral terms. Nevertheless, we can still eliminate the logarithmic singularity in $\psi_n$ by appropriately choosing integration constants that have been introduced at \emph{previous} orders (i.e.\ in the solutions for  $\psi_{n-1}$, $\psi_{n-2}, \dots$). The values of those earlier integration constants change the form of the inhomogeneity $R_n$ and, therefore, the integrals in \eqref{eq:ODEsol3}.

In order to illustrate the above considerations, we look in more detail at the cases $l=0,\dots, 4$ and $n=0,\dots,4$. In each case, we translate the condition for regularity on the integration constants into a condition for the initial data. It turns out that we have to choose data $\psi(\tau=0,\rho)$ and $\psi_{,\tau}(\tau=0,\rho)$ for which certain linear combinations of these functions and their $\rho$-derivatives vanish at $\rho=0$. The resulting conditions are listed in Table \ref{tab}. If, for a fixed $l$, all conditions up to a certain $n$ are satisfied, then all functions $\psi_0,\dots,\psi_n$ will be regular at $\tau=1$. Hence, the corresponding logarithmic terms do not appear for data subject to these conditions. Note that the regularity conditions with $l=0$ and $n=0,1,2,3$ were already derived in \cite{FrauendienerHennig2017}.

\begin{table}
%
{\small\renewcommand{\arraystretch}{1.2}
\begin{tabular}{|p{2mm}||p{2.3cm}|p{2.2cm}|p{2.5cm}|p{2.96cm}|p{3.3cm}|}
\hline
 $n$ & $l=0$ & $l=1$ & $l=2$ & $l=3$ & $l=4$ \\
\hline\hline
 $0$ & $\psi_{,\tau}=0$ & $\psi=0$ & $\psi_{,\tau}=0$ & $\psi=0$ & $\psi_{,\tau}=0$\\
\hline
 $1$ & NA 
     & $\psi_{,\tau\rho} + 10 \psi_{,\tau}\newline + \psi_{,\rho} = 0$
     & $\psi_{,\rho} + 18 \psi = 0$
     & $\psi_{,\tau\rho}+\frac{70}{3} \psi_{,\tau}\newline +\psi_{,\rho}=0$
     & $9 \psi_{,\rho}+274 \psi=0$\\
\hline
 $2$ & $\psi_{,\tau\rho}+\psi_{,\rho}\newline =0$
     & NA
     & $\psi_{,\tau\rho\rho}+2 \psi_{,\rho\rho}\newline
       +\frac43 (23 \psi_{,\tau\rho}\newline -\frac{9434}{35} \psi)=0$
     & $945 \psi_{,\rho\rho}\newline +51660 \psi_{,\rho}\newline-43264 \psi_{,\tau}=0$
     & $\psi_{,\tau\rho\rho}+2 \psi_{,\rho\rho}\newline
      +\frac85 (39 \psi_{,\tau\rho}\newline
      -\frac{452902}{693}\psi)=0$\\
\hline
 $3$ & $3 \psi_{,\rho\rho} +14 \psi_{,\rho} \newline -36 \psi=0$
     & $\psi_{,\tau\rho\rho}+2 \psi_{,\rho\rho}\newline -36 \psi_{,\tau}=0$
     & NA
     & $\frac{105}{4} (\psi_{,\tau\rho\rho\rho}
       +3 \psi_{,\rho\rho\rho}\newline
       +\frac{294}{5} \psi_{,\tau\rho\rho})\newline
       +147124 \psi_{,\tau\rho}\newline
       +(\frac{46785491}{15}\newline -173056\ln2) \psi_{,\tau}\newline =0$
     & $\frac{1}{192} \psi_{,\rho\rho\rho}
      +\frac{11}{20} \psi_{,\rho\rho}\newline
      -\frac{2312}{3465} \psi_{,\tau\rho}\newline
      -(\frac{143840317}{498960}\newline
      +\frac{73984}{693}\ln2)\psi=0$\\
\hline
 $4$ & $\psi_{,\tau\rho\rho\rho}+3 \psi_{,\rho\rho\rho}\newline
       +45 \psi_{,\tau\rho\rho} \newline
       +8 (77 \psi_{,\tau\rho}\newline +184 \psi)=0$
     & $10 \psi_{,\rho\rho\rho}\newline +123 \psi_{,\rho\rho}\newline
       +\frac{32}{5} (361\psi_{,\tau}\newline
       -90\psi_{,\rho})=0$
     & $\psi_{,\tau\rho\rho\rho}+3 \psi_{,\rho\rho\rho}\newline
      -\frac{2}{35}(3045 \psi_{,\tau\rho}\newline -34586 \psi)=0$
     & NA
     & $\frac18 \psi_{,\tau\rho\rho\rho\rho}+\frac12 \psi_{,\rho\rho\rho\rho}\newline
        +\frac{407}{35}\psi_{,\tau\rho\rho\rho}\newline
        +24(\frac{407}{5}\ln2\newline 
        -\frac{633291983}{3203200}) \psi_{,\rho\rho}\newline
        -96(\frac{407}{5}\ln^22\newline
        +\frac{1066985657107}{6054048000}\newline
        -\frac{157035547}{2882880}\ln2) \psi_{,\tau\rho}\newline
        +256(\frac{396300733}{320320}\ln^22\newline
        +\frac{17520976798265933}{1598268672000}\newline
        -\frac{452636099453}{271814400}\ln2\newline
        -\frac{10693}{126}\pi^2\newline
        -814\ln^32) \psi=0$\\
\hline 
\end{tabular}
}
\caption{Regularity conditions for the initial data that must be satisfied at $\rho=\tau=0$ for various values of $n$ and $l$. Note that ``NA'' indicates that the corresponding  function $\psi_n$ is automatically regular so that no condition needs to be imposed to achieve regularity at that order.\label{tab}}
\end{table}

\section{Simple exact solutions\label{sec:testsol}}

As a first test of the performance of the pseudospectral numerical procedure, when applied to the nonspherical wave equation, we construct simple exact ``test solutions''. Then we can  try to reconstruct them numerically. To this end, we follow the approach from \cite{FrauendienerHennig2017} and consider the conformally invariant wave equation with respect to the physical metric $\tilde g$ in (isotropic) Schwarzschild coordinates. By restricting ourselves to time-independent modes $\tilde\psi$, we can reduce the equation to an ODE. A solution $\tilde\psi$ can then be transformed to a solution $\psi$ with respect to our conformally rescaled metric \eqref{eq:conmet} using $\psi=\Theta^{-1}\tilde\psi$ (since the conformal weight in four spacetime dimensions is $-1$). 

The ODE in question can be written as
\begin{equation}
 \frac{r^2}{1-r^2}\left((1-r^2)\tilde\psi_{,r}\right)_{,r}-l(l+1)\tilde\psi=0.
\end{equation}
The general solution can be given in terms of hypergeometric functions and two integration constants. In order to achieve regularity at the cylinder (i.e.\ at $r=0$), one of the integration constants must be chosen appropriately, while the other constant can have an arbitrary value (and the particular value is not important as it simply rescales the solution by multiplication with a constant factor).
Using $r=\rho(1-\tau)$, the solution obtains a nontrivial time dependence with respect to our time coordinate $\tau$.

We explicitly construct the test solutions for $l=0,\dots, 4$. After a certain choice of the integration constants, these are
\begin{eqnarray}
 l=0:\quad && \psi=\frac{(1+r)^2}{r}\ln\frac{1+r}{1-r},\\
 l=1: && \psi=\frac{(1+r)^2}{r^2}\left((1+r^2)\ln\frac{1+r}{1-r}-2r\right),\label{eq:l=1test}\\
 l=2: && \psi=\frac{(1+r)^2}{r^3}\left((3r^4+2r^2+3)\ln\frac{1+r}{1-r}-6r(1+r^2)\right),\\
 l=3: && \psi=\frac{(1+r)^2}{r^4}\Bigg(3(1+r^2)(5r^4-2r^2+5)\ln\frac{1+r}{1-r}
         \nonumber\\
      && \qquad\qquad -2r(15r^4+14r^2+15)\Bigg),\\
 l=4: && \psi=\frac{(1+r)^2}{r^5}\Bigg((105r^8+60r^6+54r^4+60r^2+105)\ln\frac{1+r}{1-r}
         \nonumber\\
      && \qquad\qquad -10r(1+r^2)(21r^4-2r^2+21)\Bigg).
\end{eqnarray}
Note that the terms in large parentheses all vanish sufficiently quickly as $r\to0$ and  compensate the singularities of the first factors at $r=0$. In fact, the solutions are analytic in our domain $0\le\tau\le1$, $0\le\rho\le\rho_\mathrm{max}<1$.

Now we prescribe initial data for these test solutions at $\tau=0$ and see how accurately we can recover the corresponding solutions numerically. \JH{An} interesting numerical detail is as follows. In the limit $\tau\to1$, the wave equation \eqref{eq:modes} becomes
\begin{equation}\label{eq:atScri}
 2(\rho\psi_{,\tau\rho}-\psi_{,\tau})-l(l+1)\psi=0.
\end{equation}
In the spherically symmetric case $l=0$, this is an intrinsic equation for $\psi_{,\tau}$ with solution $\psi_{,\tau}=c\rho$. The integration constant $c$, however, cannot be determined from the wave equation at $\Scri^+$. Consequently, we observed in \cite{FrauendienerHennig2017} that we cannot just impose the wave equation itself at all grid points at $\Scri^+$; an extra condition is required that fixes the integration constant. This can be done by prescribing the exact value of the mixed derivative $\psi_{,\tau\rho}$ at the single point $\rho=0$, $\tau=1$, which equals $c$. This value can easily be obtained from the analysis at the cylinder.

Is such an extra condition also required for nonspherical waves, i.e.\ for $l>0$? In that case, \eqref{eq:atScri} is no longer an intrinsic equation for $\psi_{,\tau}$, due to the additional $l(l+1)$-term. Hence it should be possible to impose the wave equation at all grid points. On the other hand, if extra information about the solution to a PDE problem is available and can explicitly be enforced in the numerical code, this may improve the numerical accuracy. \JH{Therefore, it is worth comparing} the numerical performance with and without an extra condition at $\rho=0$, $\tau=1$. As an extra condition, we can again give the exact value for $\psi_{,\tau\rho}$. For example, in the case $l=1$, this condition reads
\begin{equation}\label{eq:extra}
 l=1:\quad \psi_{,\tau\rho}\big|_{\tau=1,\rho=0}
     = - 4(1-\ln 2) \psi_{,\tau} - \psi_{,\rho}\big|_{\tau=0,\rho=0},
\end{equation}
where the right-hand side can be computed from the initial data at $\tau=0$.

\begin{figure}\centering
 \includegraphics[width=8cm]{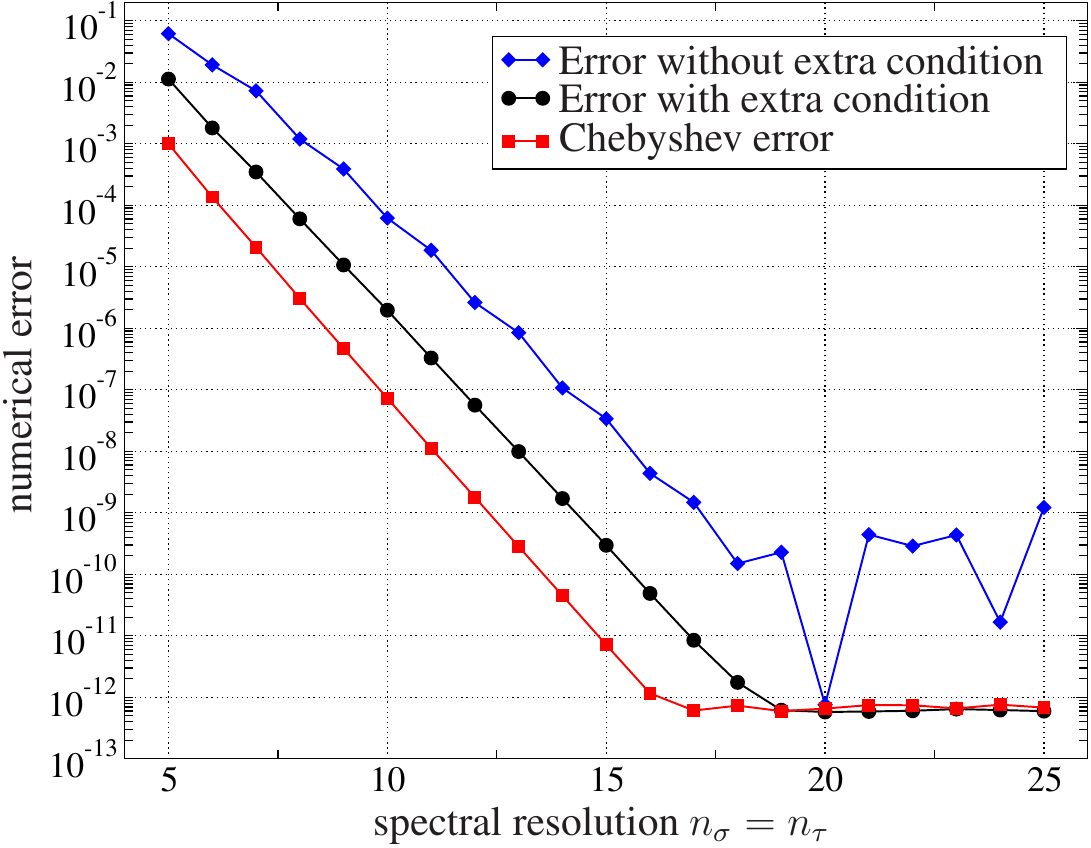}
 \caption{The error in the numerical reconstruction of the $l=1$ test solution \eqref{eq:l=1test}  with parameter $\rho_\mathrm{max}=0.5$. The resolutions (i.e.\ number of collocation points) in spatial and time direction are denoted with $n_\sigma$ and $n_\tau$, respectively. In all examples discussed in this article we always choose $n_\sigma=n_\tau$, but other choices are of course possible as well.\label{fig:extra}}
\end{figure}
The corresponding convergence plot is shown in Fig.~\ref{fig:extra}. As a measure \JH{of} the numerical error, we compute the largest absolute difference between the exact and numerical solutions at $100\times100$ equidistant grid points. As expected, we find that the solution can be obtained with good accuracy by just imposing the wave equation at all grid points. However, the accuracy significantly further improves if we use the extra condition \eqref{eq:extra}. For comparison, \JH{Fig.~\ref{fig:extra}} also shows the Chebyshev interpolation error, which is the largest difference between the exact solution and a Chebyshev approximation to this solution (without solving any differential equation). We see that the extra condition allows the numerical accuracy to reach this ``theoretical threshold.'' 

For the test solutions with other values of $l$, we qualitatively obtain the same picture. Hence, from now on, we will always impose the extra conditions at $\rho=0$, $\tau=1$, in order to optimise \JH{numerical accuracy.}

Another important feature of the convergence plot in Fig.~\ref{fig:extra} (and the plots obtained for other $l$ values) is that the error approximately follows a straight line \JH{on a logarithmic scale}, until the accuracy is close to machine accuracy. This is exactly what is expected for sufficiently regular (namely $C^\infty$ or analytic) solutions like our test solutions.

\section{Logarithmic singularities\label{sec:logsing}}

We have seen that the numerical method enables us to very accurately reconstruct the analytic test solutions. For general initial data, however, we know that logarithmic singularities form on the cylinder. Moreover, we should expect that these singularities also ``spread'' to $\Scri^+$. Whether this is true will be investigated in the second half of this section.

Similarly to our considerations in \cite{FrauendienerHennig2017}, we start by constructing families of initial data that have their leading singularities on the cylinder at a prescribed order. For the first few values of $l$, we choose the following data at $\tau=0$, which depend on two free parameters $\alpha$, $\beta$,
\begin{eqnarray}\label{eq:ab1}
 \fl l=1:&& \psi = \rho-\frac{288+\beta}{2065}\sin(5\rho),\quad
  \psi_{,\tau} = \frac{\alpha}{10}\cos(2\rho)-\left(\alpha+\frac{125-\beta}{413}\right)\rho,\\
 \fl l=2: && \psi=\rho-\frac15\sin(5\rho),\nonumber\\
 \fl &&\psi_{,\tau}=\frac{3}{184}\left(\frac12(87\alpha+\beta+1150)\sin(2\rho)-(89\alpha+\beta+1150)\rho\right),\\
 \fl l=3: && \psi = \rho+\big[865280(10816\alpha-12915)\ln2+17307471168\alpha -54080\beta
    \nonumber\\
  \fl&&\qquad -28622700215\big]/[25(2235018240\ln2+5745834043)]\sin(5\rho),\nonumber\\
 \fl  && \psi_{,\tau}=
 \big[15(-6450348736\alpha-51660\beta+1017056256)\cos(2\rho)\nonumber\\
 \label{eq:ab3}
 \fl &&\qquad -112306421760(\ln2)\alpha\rho+(2049932403584\alpha+18729960\beta\nonumber\\
 \fl &&\qquad -36874608750)\rho\big]/[134101094400\ln2+344750042580].
\end{eqnarray}
These families are constructed, using the regularity conditions in Table~\ref{tab}, such that for any choice of the parameters, the functions $\psi_0$ and $\psi_1$ in the expansion \eqref{eq:expan} are free of logarithmic terms.\footnote{This is the minimum requirement to ensure that the wave function $\psi$ and the derivatives up to second order, i.e.\ the quantities that explicitly appear in the wave equation, are bounded in our numerical domain. Otherwise, our numerical method fails and cannot solve the problem. However, we will later consider the equation in different coordinates that allow us to reconstruct even such ``very singular'' solutions, see Sec.~\ref{sec:reg} below.} Moreover, for $\alpha\neq0$, the next possible logarithmic singularity is present at the cylinder (for $l=1$ a singularity in the function $\psi_3$, since $\psi_2$ is then automatically regular; and a singularity in $\psi_2$ for $l=2, 3$). This singularity disappears for $\alpha=0$. For $\beta=0$, the following logarithmic term vanishes as well.

\begin{figure}\centering
 \includegraphics[width=7.6cm]{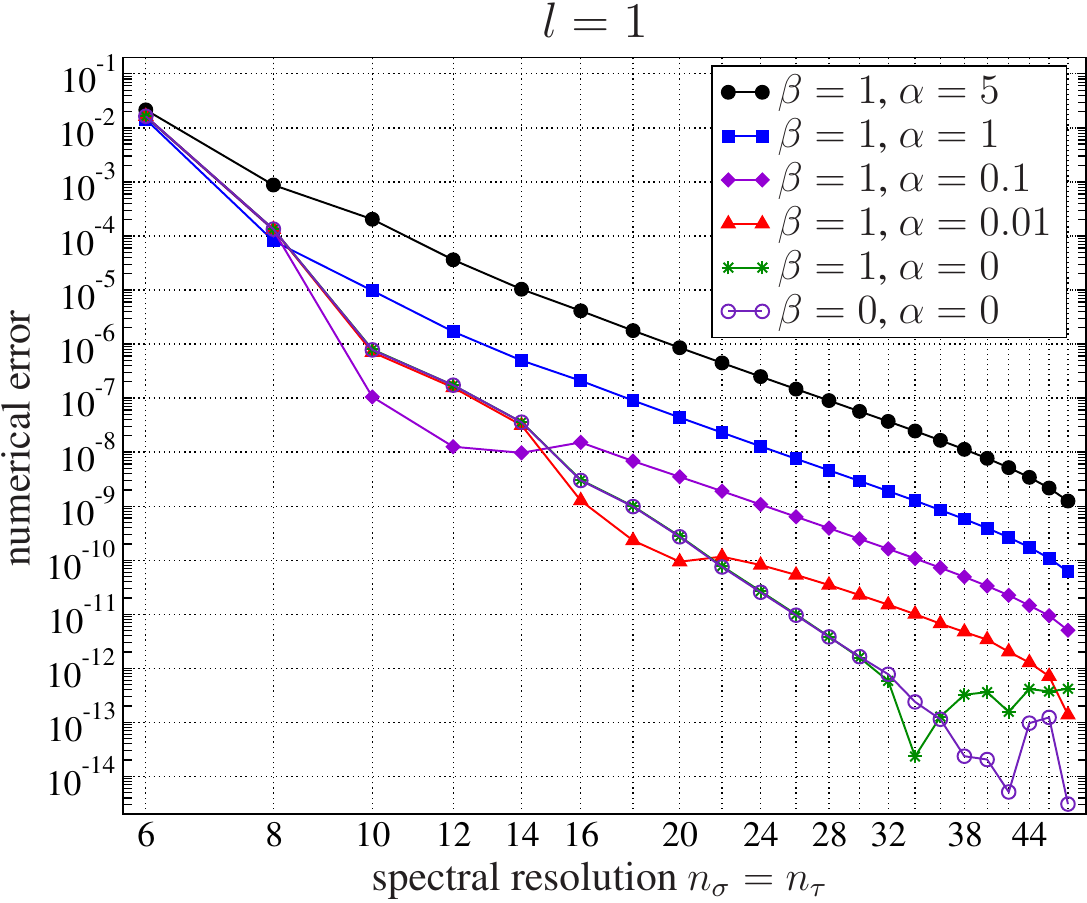}\quad
 \includegraphics[width=7.6cm]{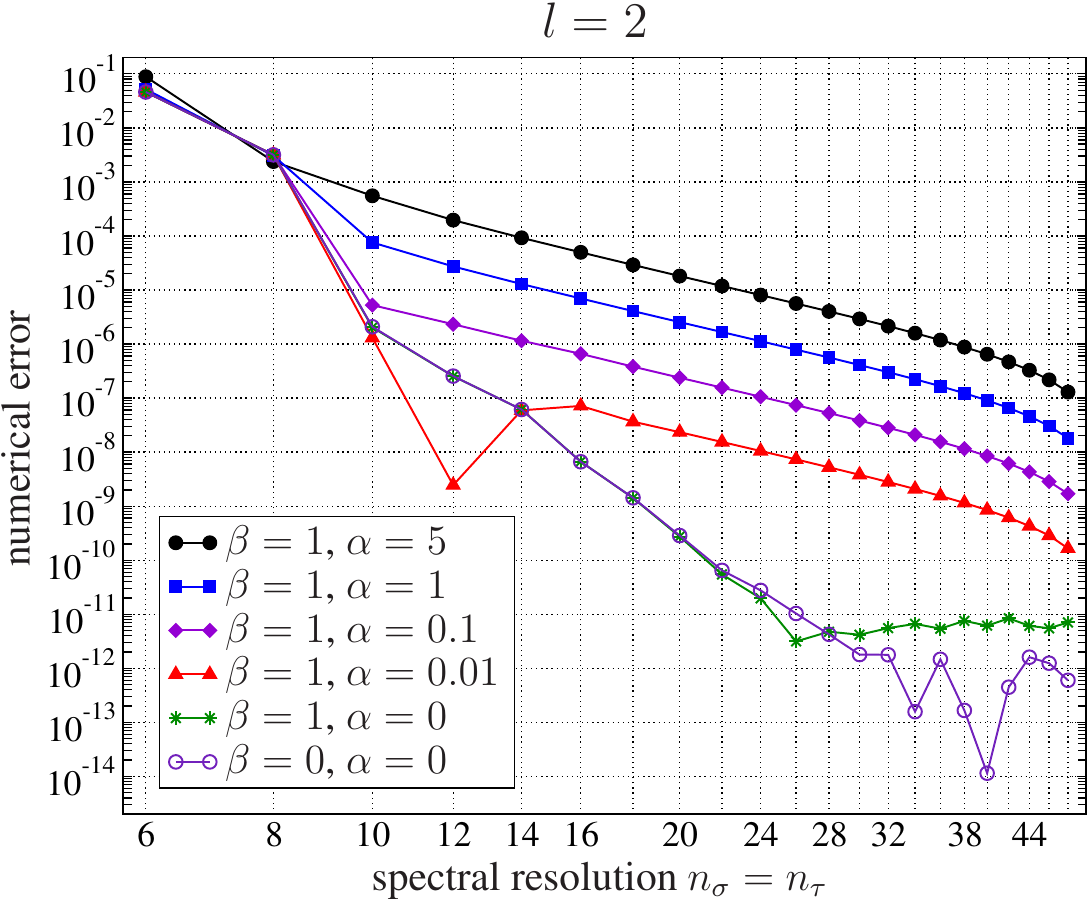}\\
 \includegraphics[width=7.6cm]{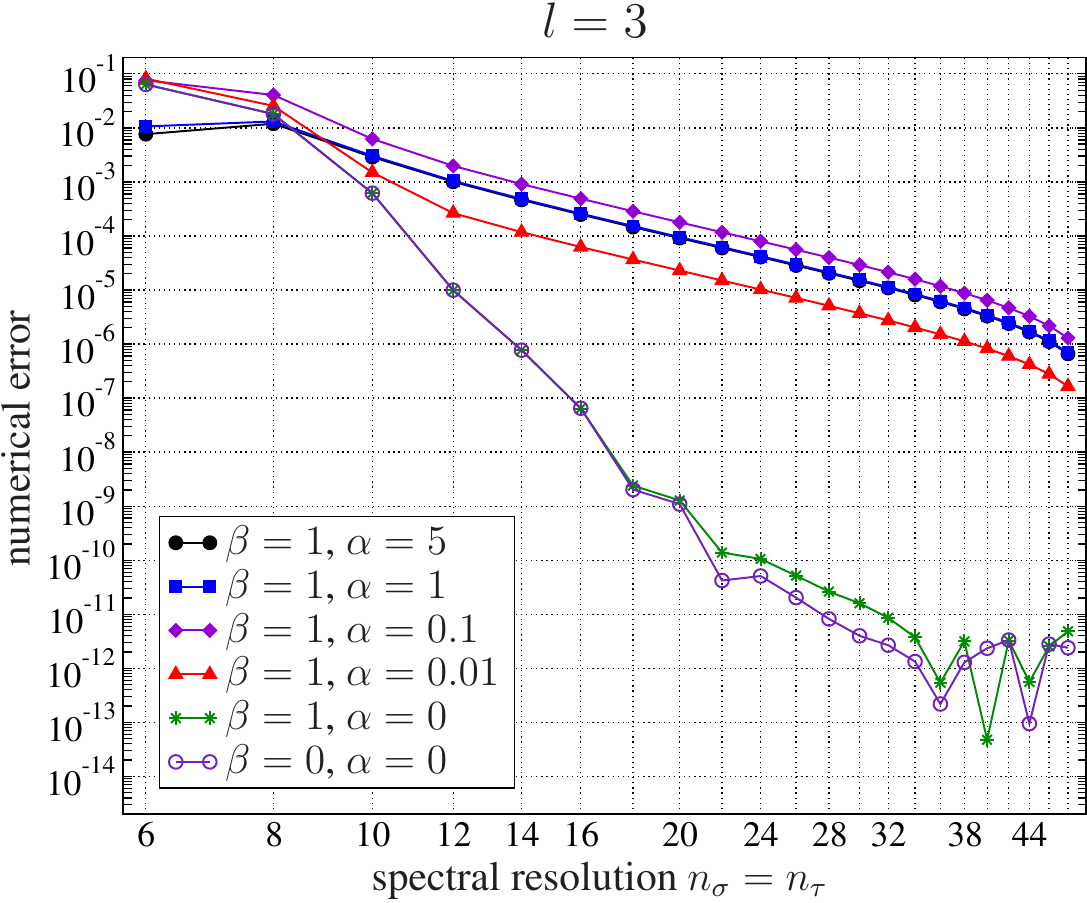}
 \caption{Convergence plots for the families of initial data \eqref{eq:ab1}--\eqref{eq:ab3}  for various parameter values in the cases $l=1, 2, 3$ (and with coordinate parameter $\rho_\mathrm{max}=0.5$).\label{fig:alphabeta}}
\end{figure}
We solve the corresponding initial value problems for different values of the parameters, in order to compare examples in which the leading logarithmic terms appear at different orders and have coefficients of different magnitudes. The numerical accuracy is then tested by estimating the relative error of the function value $\psi$ at $\tau=1$, $\rho=\rho_\mathrm{max}$ (i.e.\ a point at $\Scri^+$), which is done by comparing the function values as obtained with different resolutions with the value corresponding to the highest resolution of $n_\sigma=n_\tau=52$. The resulting convergence plots are shown in Fig.~\ref{fig:alphabeta}. 

We observe that the accuracy generally increases for decreasing values of the parameters $\alpha$ and $\beta$, since the coefficients of the leading singular terms then become smaller or even vanish (if $\alpha=0$ or $\beta=0$ are chosen). Clearly, more regular solutions can be resolved with greater accuracy. Most importantly, the curves roughly follow straight lines (for sufficiently large resolutions) \JH{on} a \emph{double}-logarithmic scale, corresponding to \emph{algebraic} rather than exponential convergence. This shows that these solutions only have finite differentiability $C^k$ instead of being $C^\infty$. In such cases the error is indeed expected to behave like a negative power of the resolution, which is exactly the observed algebraic convergence. Evidently, the logarithmic singularities have a strong influence on the numerical results. In particular, if logarithmic terms are already present at low orders, then much larger resolution \JH{is} required --- compared to the previously discussed analytic test solutions --- to obtain a reasonable accuracy. This is qualitatively the same behaviour that we already found for $l=0$ in \cite{FrauendienerHennig2017}.

An important question is whether the singularities are all concentrated at $I^+$ (i.e.\ at $\tau=1$, $\rho=0$), where the cylinder and $\Scri^+$ approach each other, or whether the solutions also have a limited differentiability as $\tau\to1$ for $\rho>0$. This will be investigated in the following.

In order to study the regularity of a solution $\psi$ as $\Scri^+$ is approached, we  consider $\psi(\tau,\rho)$ as a 1-dimensional function of $\tau$ for some fixed $\rho$. From the numerical solution we read off the function values of $\psi$ at the collocation points on this line (via Chebyshev interpolation). Then we compute the Chebyshev approximation of this 1-dimensional function. The fall-off behaviour of the Chebyshev coefficients $c_k$ depends on the regularity of the function: for a $C^\infty$ function, the coefficients fall off exponentially, while for a $C^k$ function, they fall off algebraically. In particular, if a function behaves like $(1-\tau)^s\ln(1-\tau)$ (plus weaker singular and plus regular terms) near $\tau=1$ for some integer $s$ and is $C^\infty$ for $0\le\tau<1$, then, for sufficiently large $k$, we expect\footnote{This can be shown with repeated partial integration of the integral formula for the Chebyshev coefficients.} $c_k\sim k^{-(2s+1)}$. If we plot $\log_{10}|c_k|$ versus $\log_{10}k$ in a double-logarithmic diagram, then we should roughly find a straight line with slope $-(2s+1)$, at least in an appropriate range of $k$ values.

\begin{figure}\centering
 \includegraphics[width=7.6cm]{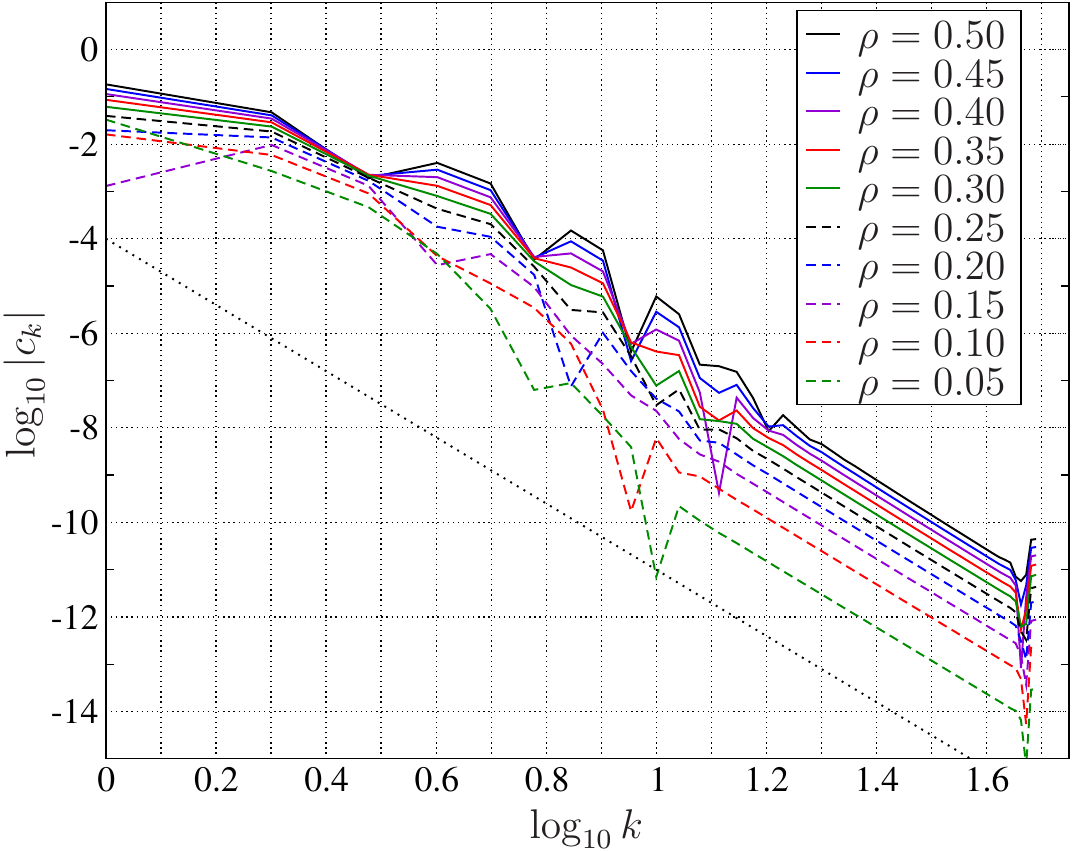}\quad
 \includegraphics[width=7.6cm]{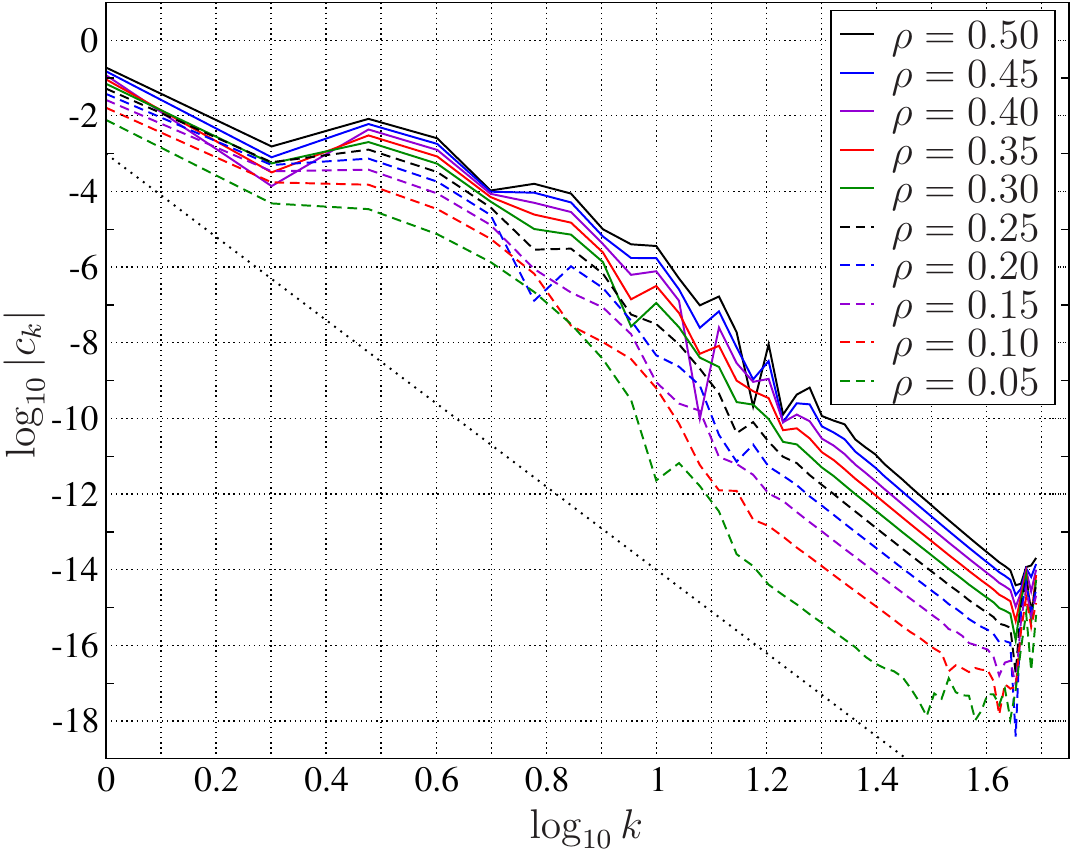}
 \caption{Fall-off of the Chebyshev coefficients $c_k$ of $\psi(\tau,\rho)$, considered as a 1-dimensional function of $\tau$ for different fixed values of $\rho$. The solutions correspond to the case $l=1$ with initial data \eqref{eq:ab1} with parameters $\alpha=\beta=1$ (left panel) and $\alpha=0$, $\beta=1$ (right panel). For comparison, dotted straight lines with slopes $-7$ (left panel) and $-11$ (right panel) are shown. Numerical parameters: $\rho_\mathrm{max}=0.5$, $n_\sigma=n_\tau=50$.\label{fig:sing}}
\end{figure}
Figure~\ref{fig:sing} illustrates the observed fall-off for two solutions, corresponding to different initial data in the case $l=1$. In the first case (as shown in the left panel), we approximately find straight lines with slope $-7$ for sufficiently large $k$ (but $k$ not too close to the maximum\footnote{If $k$ is close to the number of collocation points (here $50$, i.e.\ for $\log_{10}k$ close to $1.7$), then we see deviations from the linear behaviour. This comes from a combination of two effects. The first is due to numerical errors, which lead to the strongest deviations from the straight line for the smallest Chebyshev coefficients\JH{, with magnitude of} the order of the numerical error. The second effect comes from the difference between the coefficients of an infinite Chebyshev series (which strictly have the algebraic fall-off discussed above) and the coefficients in Chebyshev interpolation with a finite number of coefficients (as considered here), which becomes most significant if $k$ is close to the maximum, see~\cite{Boyd}.}). For different values of $\rho$, the magnitudes of the Chebyshev coefficients change (and they tend to become smaller as $\rho\to0$, in accordance with regularity of $\psi$ on the cylinder $\rho=0$, where only $\rho$-derivatives of $\psi$ have logarithmic terms, but not $\psi$ itself), but they approach zero at the same rate as the resolution increases. This is consistent with a logarithmic singularity of the form $(1-\tau)^3\ln(1-\tau)$. In the second case (right panel), we observe a slope of about $-11$, corresponding to a logarithmic term $(1-\tau)^5\ln(1-\tau)$.

In addition to the examples from Fig.~\ref{fig:sing}, we have investigated the fall-off for many other solutions. In particular, for all slopes in the set $\{-5,-7,-9,-11,-13\}$, we found examples for which the Chebyshev coefficients have a fall-off corresponding to straight lines with these slopes. This is consistent with leading logarithmic terms of the form $(1-\tau)^s\ln(1-\tau)$ with $s=2,3,4,5,6$. Of course, these observations are no rigorous proof of exactly that form of singularity, as other types of singularities could lead to a similar fall-off behaviour. But inspired \JH{by} our analytical considerations on the cylinder, we strongly expect to find exactly those logarithmic singularities. Depending on the initial data, the leading-order singularity can appear at different orders, as our examples show. 

One important point that still needs some attention is to verify whether the singularities are indeed concentrated at $\Scri^+$. In other words, we have to make sure that along the considered lines $\rho=\mathrm{constant}$ no singularities occur before $\tau=1$ is reached. This is already clear from an analytical point of view, since solutions to the \emph{linear} wave equation on a Schwarzschild background corresponding to smooth initial data are smooth. But we want to verify that the same shows in our numerics.
To this end, we can use the same approach as before, but this time applied to lines $\tau=\mathrm{constant}$. If $\psi$ is regular on such lines for $\tau<1$, then the Chebyshev coefficients of $\psi$ (now taken to be a function of $\rho$) should fall off exponentially, rather than algebraically. Corresponding numerical experiments indeed verify that this is the case.

\begin{figure}\centering
 \includegraphics[width=8cm]{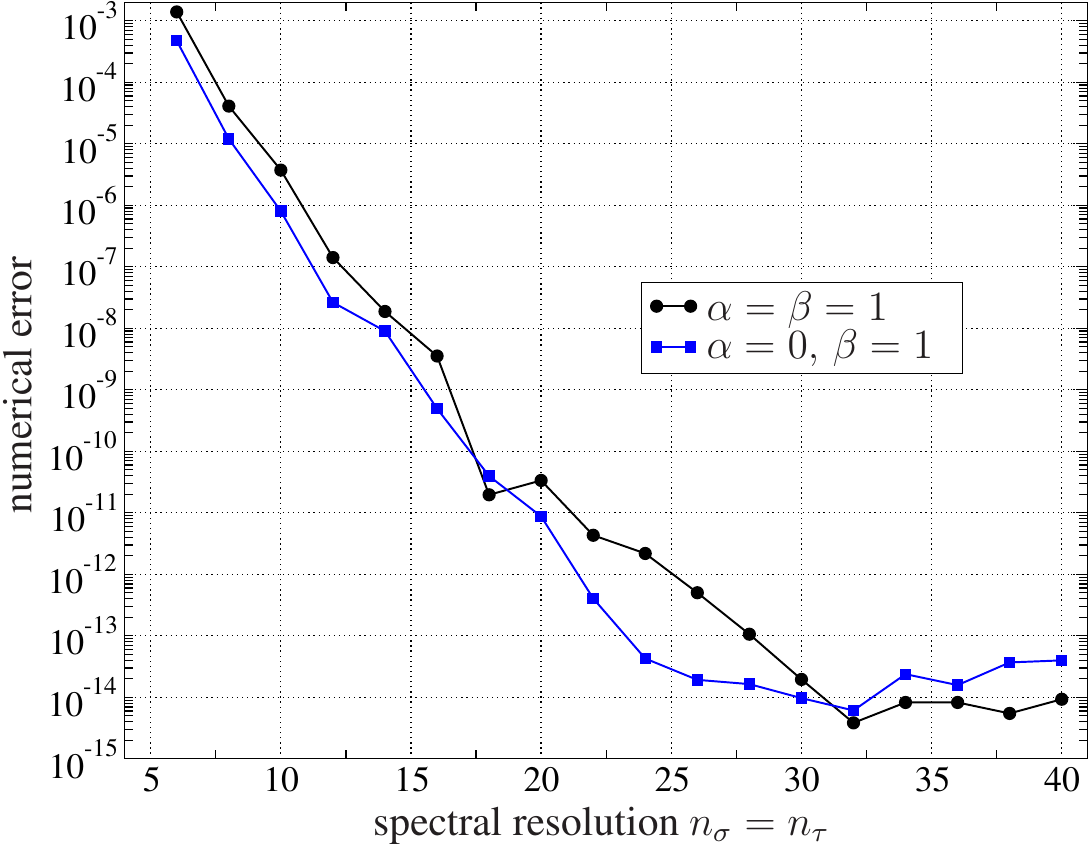}
 \caption{Convergence plot for two solutions corresponding to the same sets of initial data as considered in Fig.~\ref{fig:sing}, but this time computed in the smaller time interval $0\le\tau\le0.95$. The relative error quickly reaches values below $10^{-13}$, and the observed exponential convergence shows that this domain is free of singularities.\label{fig:smaller}}
\end{figure}

Another way to show that the region $\tau<1$ is free of singularities is to solve the wave equation in the restricted domain $0\le\rho\le\rho_\mathrm{max}$, $0\le\tau\le\tau_\mathrm{max}$ for some $\tau_\mathrm{max}<1$ but close to $1$. If there are no singularities in this domain, then the measured numerical error should decrease exponentially. This is exactly what we observe, see Fig.~\ref{fig:smaller}.

We therefore conclude that singularities occur at $\tau=1$ for all $\rho$-values, i.e.\ not just at the cylinder but at $\Scri^+$ as well, while the solutions are $C^\infty$ in the region $\tau<1$. The observed fall-off of the Chebyshev coefficients is consistent with (\JH{though not rigorous proof of) the} presence of logarithmic singularities at certain leading orders. The exact order depends on the particular solution, i.e.\ on the choice of initial data.

\section{Regularisation of (too) singular solutions\label{sec:reg}}

The previous numerical calculations show that the technique of fully pseudospectral time evolution can successfully be applied to nonspherical modes of the Schwarzschild wave equation. We also observe that the numerical accuracy depends on the lowest order at which logarithmic singularities occur. If this order is not ``too low'', then highly accurate solutions are possible. Otherwise, however, only a moderate accuracy can be expected. If the solutions are singular enough so that even derivatives of second order or lower diverge as $\tau\to 1$, then quantities appearing in the equation itself are unbounded. In this case no reliable degree of accuracy should be expected at all.

This raises the question as to whether the formulation of the problem can be changed to allow for more accurate numerical solutions and to make even initial value problems with  very singular solutions tractable. Based on our observation that logarithmic singularities appear at $\Scri^+$, a possible approach is to reformulate the wave equation in terms of different coordinates such that $\psi$ as a function of these new coordinates is more regular. This will be described in the following.

We use a generalisation of a coordinate transformation that was applied in \cite{KalischAnsorg} to construct black string solutions in higher dimensions.\footnote{Note that the problem studied in \cite{KalischAnsorg} required the solution of elliptic equations to construct a stationary equilibrium configuration. At a certain spatial boundary, logarithmic singularities were observed and regularised with a spatial coordinate transformation. Here we apply a generalisation of the same transformation to the \emph{time} coordinate, which turns out to work as well. An important new ingredient provided here is the free parameter $w$ in \eqref{eq:trans}.} To this end, we replace our time coordinate $\tau$ with a new coordinate $t$ by setting
\begin{equation}\label{eq:trans}
 \tau=1-\ee^{-\frac{wt}{1-t}},
\end{equation}
where $w>0$ is a parameter. Suitable values for $w$ have to be identified with numerical experiments, and we will see that an appropriate choice of $w$ is crucial for \JH{numerical accuracy.} Note that this transformation maps the interval $[0,1]$ to itself. Moreover, a term of the form $(1-\tau)^{s_1}\ln^{s_2}(1-\tau)$ with $s_1,s_2>0$ 
(which is a slightly more general logarithmic singularity with respect to $\tau$ than discussed before), becomes $\ee^{-\frac{ws_1t}{1-t}}\left(\frac{-wt}{1-t}\right)^{s_2}$ in terms of $t$. The singularity of the second factor is compensated by the first factor, which exponentially goes to zero as $t\to1$. Hence a function with singularities of this type, together with all of its $t$-derivatives, will be regular in terms of $t$. Consequently, we can transform (particular) $C^k$ functions into $C^\infty$ functions. 

In order to test whether this procedure improves the numerical accuracy, we reformulate the wave equation in terms of $t$. The result is
\begin{eqnarray}
  \frac{1}{w}\left[2-(1-\tau)A\right](1-t)^2\psi_{,tt}\nonumber\\
  -\left[\frac{1}{w}\left(2-(1-\tau)A\right)(2-w-2t)+2\left(1-(1-\tau)A\frac{1-2r}{1-r^2}\right)\right]\psi_{,t}\nonumber\\
  +2\rho\psi_{,t\rho}+\frac{w(1-\tau)}{(1-t)^2}A\,l(l+1)\psi
 +\frac{4rAw(1-\tau)}{(1+r)^2(1-t)^2}\psi=0,
\end{eqnarray}
with $A$ and $r$ as defined in \eqref{eq:defA}, and where $\tau$ is considered to be a function of $t$. In the limit $t\to1$, this equation reduces to $\psi_{,t\rho}=0$, which we impose as condition at the upper boundary $t=1$ --- with exception of the point $t=1$, $\rho=0$, where we use the condition $\psi_{,t}=0$ instead. (Due to the structure of our new time coordinate, \emph{all} derivatives of $\psi$ that contain at least one $t$-derivative must vanish at $\Scri^+$, so it is clear that these conditions must hold.)

In the following, we solve the reformulated wave equation for solutions with different degrees of regularity, in order to test in which cases the new coordinates improve the numerical performance, and to find appropriate values for the parameter $w$.
\begin{figure}\centering
 \includegraphics[width=7.6cm]{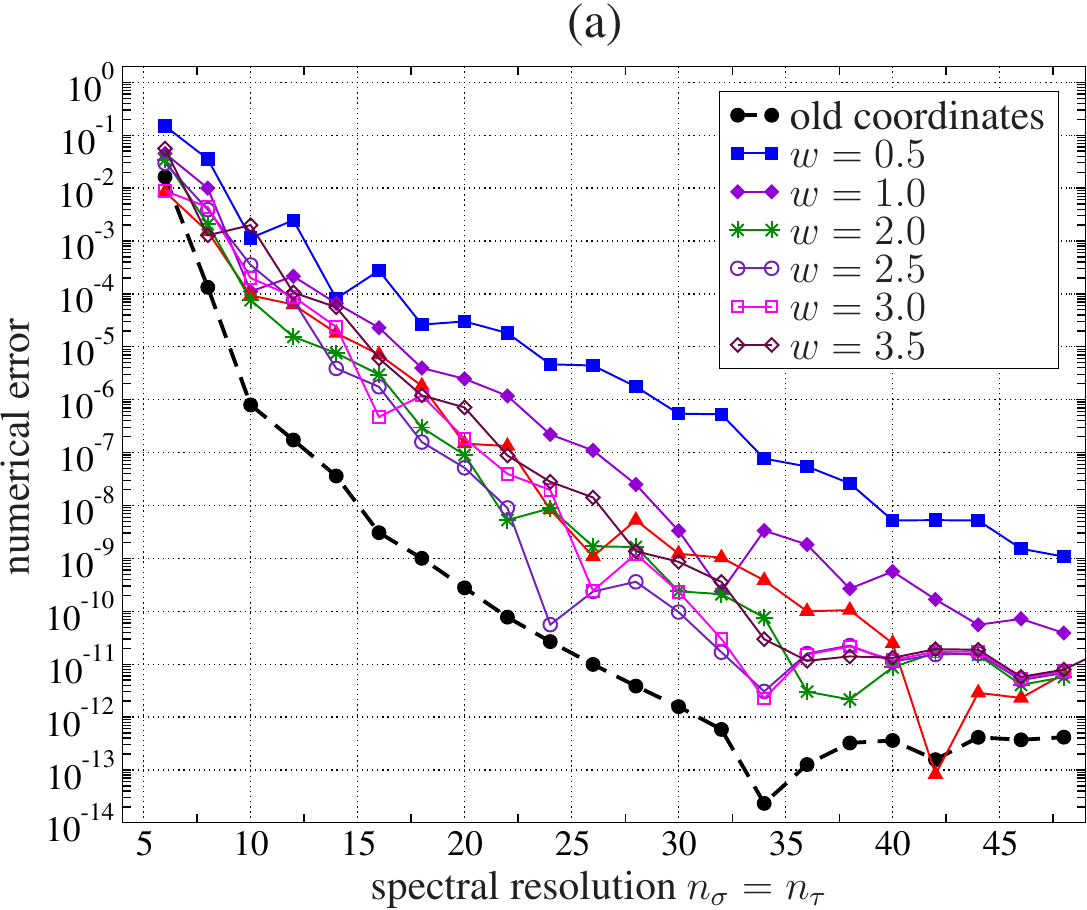}\quad
 \includegraphics[width=7.6cm]{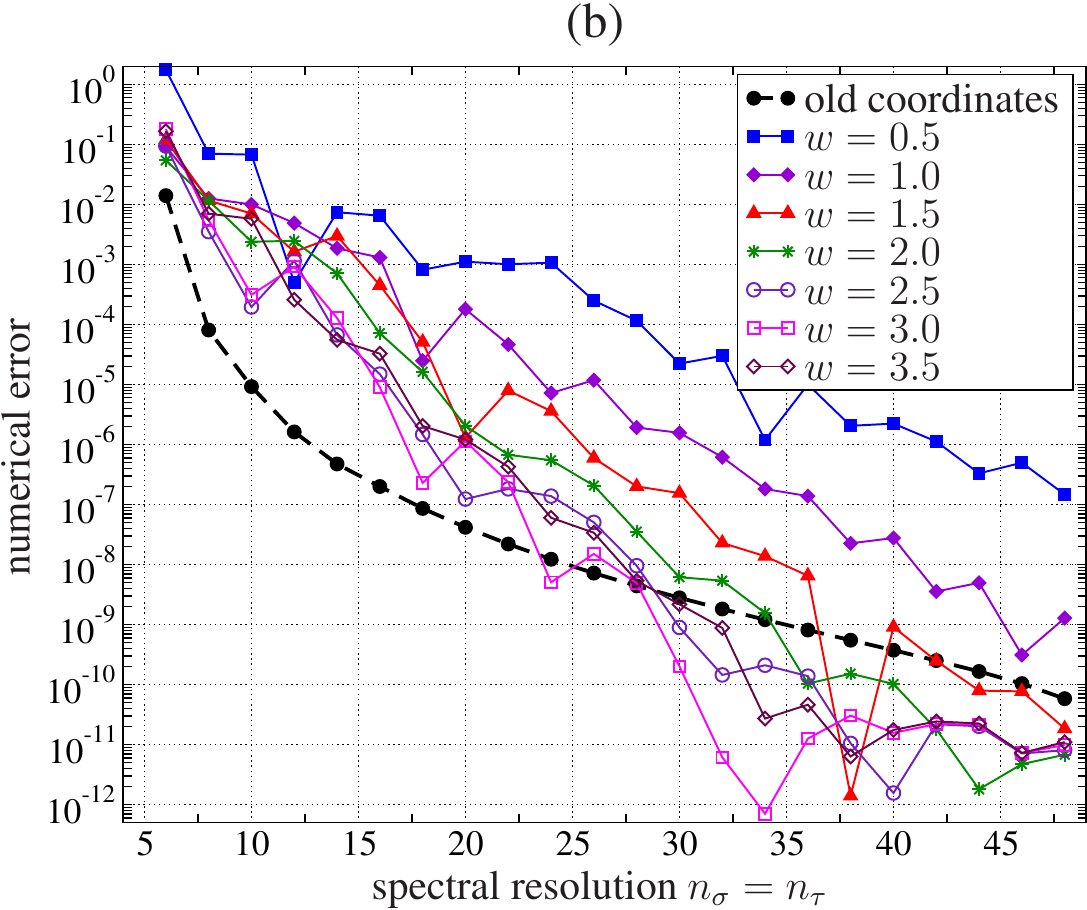}\\[1ex]
 \includegraphics[width=7.6cm]{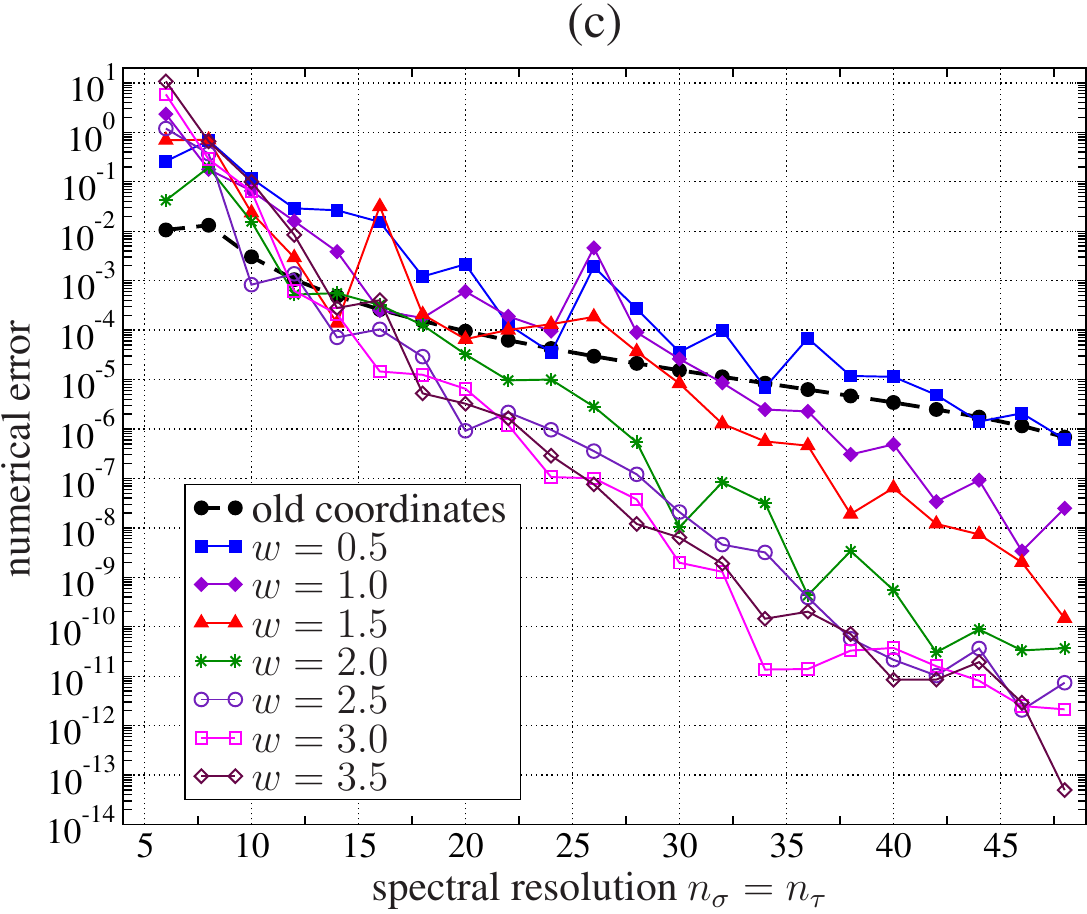}\quad
 \includegraphics[width=7.6cm]{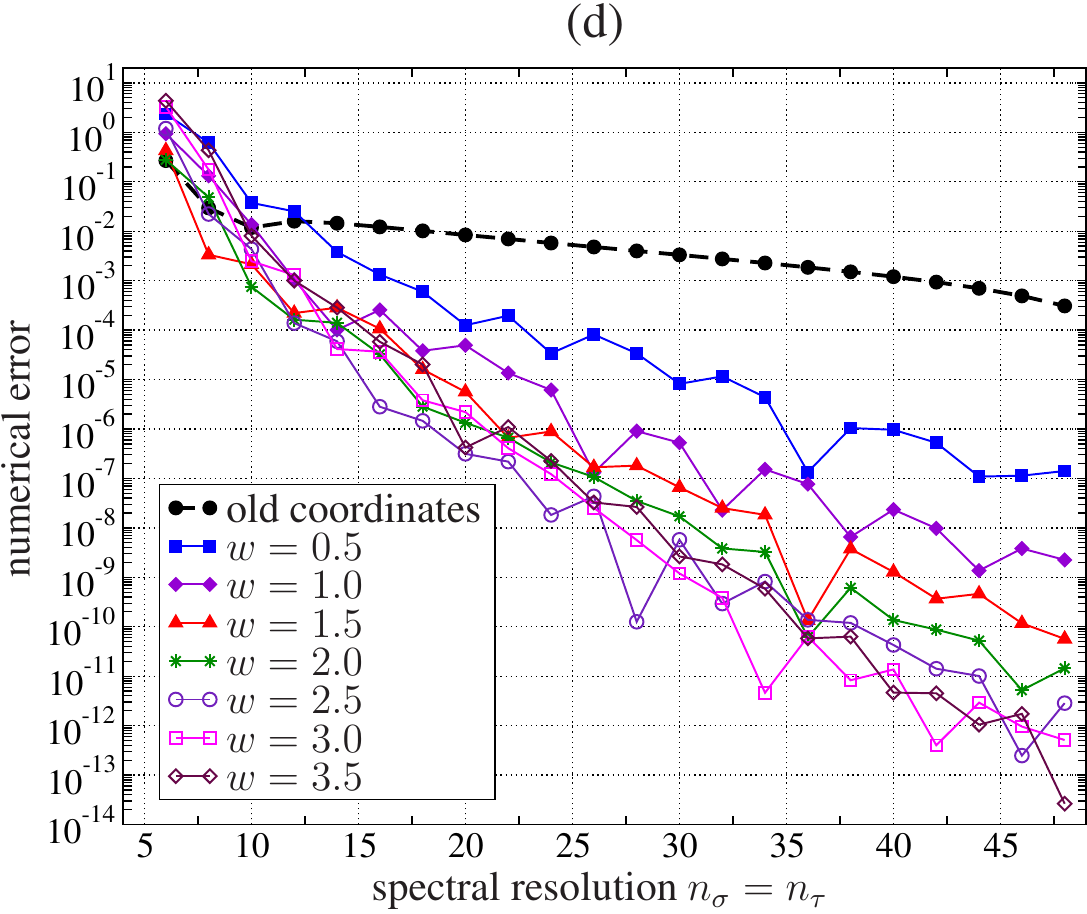}
 \caption{Comparison of the numerical accuracy that can be achieved with the new time coordinate $t$ for different choices of the parameter $w$, and the accuracy that is obtained with the old time coordinate $\tau$. The four examples (a)--(d) correspond to different initial data and  different values of the parameter $l$  as described in the main text.\label{fig:exp}}
\end{figure}
The convergence plots for four different numerical examples are shown in Fig.~\ref{fig:exp}. In the following we explain and discuss these examples.

In Fig.~\ref{fig:exp}a we consider the case $l=1$ and choose initial data \eqref{eq:ab1} with $\alpha=0$, $\beta=1$. For this choice of parameters, the regularity conditions for $\psi_0,\dots,\psi_3$ in Table~\ref{tab} are all satisfied. Consequently, the first logarithmic term appears at sufficiently high order to allow for highly accurate numerical results in the old coordinates (with relative error below $10^{-12}$). When we solve the same initial value problem with the new coordinates, then we find that we lose about one order of magnitude in accuracy for the best value of $w$, and even more otherwise. Hence the new coordinates are not a good choice to resolve the ``relatively well-behaved'' function $\psi$ in this example.

The convergence plot for the same $l=1$ family of initial data as before, but now with parameter values $\alpha=5$, $\beta=1$, is shown in Fig.~\ref{fig:exp}b. Since $\alpha\neq0$ here, singularities already appear at a lower order than in the previous example. This reduces the accuracy that can be achieved with the old coordinates by about two orders of magnitude (with a relative error of approximately $10^{-10}$ for the largest considered resolution). For an appropriate choice of $w$, the new coordinates can achieve a better accuracy (with error around $10^{-11}$), and this is obtained for an even smaller number of collocation points. Hence the new coordinates provide a certain advantage  both in terms of efficiency and accuracy in this example.

Next, in Fig.~\ref{fig:exp}c, we consider the case $l=3$ with initial data \eqref{eq:ab3}, and we choose $\alpha=5$, $\beta=1$. Using the old coordinates, the numerical error is $10^{-6}$ in the optimal case, which is still reasonably accurate, but far from the high accuracy that one usually expects from (pseudo) spectral methods. With the new coordinates, however, we can reduce the error to below $10^{-11}$ for appropriate values of $w$, i.e.\ we can improve the accuracy by five orders of magnitude.

Finally, for the example in Fig.~\ref{fig:exp}d, we choose $l=2$ and consider the initial data
\begin{equation}
 \tau=t=0:\quad \psi=\rho(\rho+1),\quad \psi_{,\tau}=0.
\end{equation}
These data satisfy the regularity condition for $\psi_0$, but not the one for $\psi_1$ in Table~\ref{tab}. This means that the mixed derivative $\psi_{,\tau\rho}$ will diverge along the cylinder in the limit $\tau\to1$. Similar singularities are expected at $\Scri^+$. With these properties, the solution is too singular to be recovered in terms of the original coordinates: the smallest relative error is only slightly below $10^{-3}$, which is not an acceptable level of accuracy to take this as a reliable numerical result (at least from the point of view of someone whose work goes beyond low-order finite difference methods). On the other hand, if we solve the same problem with the new coordinates, we can achieve an error below $10^{-12}$. Hence the new coordinates are essential to solve initial value problems that lead to very singular solutions, in which case the accuracy is immensely improved.

In summary, we observe that the original coordinates should be preferred whenever lower-order singularities are eliminated by appropriately restricting the initial data. However, when more singular solutions are considered, the new coordinates are extremely valuable in terms of the efficiency and accuracy that can be achieved. In particular, some solutions are too singular for a reasonable treatment with the old coordinates, but can be described very accurately with the new coordinates. 

Finally, in all considered examples, we observed that the parameter $w$ in the new coordinates should be chosen somewhere near $w=3$ for the best results.

\section{Discussion\label{sec:discussion}}
As a toy model for physical equations like the conformal field equations and, in particular, in order to understand the behaviour of solutions near spacelike infinity, the conformally invariant wave equation on a Schwarzschild background was investigated. We did not impose any symmetry requirements on the solutions $\psi$ and hence extended the earlier discussion of spherically symmetric solutions \cite{FrauendienerHennig2017} to the general case. With a decomposition into spherical harmonics $Y_{lm}$, the $3+1$ dimensional problem can be reduced to $1+1$ dimensional equations, which turn out to depend on $l$ but not on $m$. These equations \JH{are} degenerate on $I$, the cylinder representation of spacelike infinity $i^0$, which allows \JH{us} to derive ordinary differential equations for $\psi$ and its spatial derivatives at $I$. \JH{An analysis} of these ODEs revealed that these functions generally suffer from logarithmic singularities at infinitely many orders. However, the solutions to the ODEs also allowed us to relate the singularities at time $\tau=1$ to the initial data at $\tau=0$. In particular, we can choose initial data subject to appropriate regularity conditions, such that the corresponding time-evolution of these data does not contain the first singular terms.

We also demonstrated that the fully pseudospectral time-evolution scheme can successfully be applied to the problem in question, and that highly accurate solutions can be obtained. An analysis of the numerical solutions showed that singularities are not restricted to the cylinder $I$, but that they generically also appear everywhere on future null infinity $\Scri^+$. The observed fall-off behaviour of certain Chebyshev coefficients is consistent with logarithmic singularities at various orders.

\JH{In addition}, we have studied how the order at which the first singularities appear influences the accuracy of the numerical results. If singularities are not present at very low orders, then highly accurate solutions close to machine accuracy are directly obtained. For ``more singular'' solutions, one can significantly improve the accuracy with an adapted new time coordinate. The corresponding coordinate transformation \JH{ensures} that the solutions are infinitely differentiable with respect to the new time coordinate, even though they have logarithmic singularities with respect to the old coordinate. After this modification, highly accurate results are again obtained.

As discussed, our considerations show that solutions to the conformally invariant wave equation generally have \JH{limited regularity}, both on the cylinder $I$ and at $\Scri^+$. The same is expected for solutions to more realistic equations. Nevertheless, \JH{as with the} present investigations, we predict that the corresponding initial value problems can be tackled with the fully pseudospectral method --- provided that one pays careful attention to the singularity structure and uses suitable formulations of the problems.

\section*{Acknowledgments}
 We would like to thank \JH{Matt Parry} for commenting on the manuscript. Part of this research was supported by the University of Otago through a Division of Sciences Strategic Seeding Grant.
 


\end{document}